\documentclass[journal]{IEEEtran}
\usepackage{cite,url,subfigure,epsfig,graphicx}
\usepackage{tipa}
\usepackage{makecell}
\usepackage{multirow}
\usepackage{bbm}
\usepackage{mathrsfs}
\usepackage{array}
\usepackage{amsfonts}
\usepackage{cite,url,subfigure,epsfig,graphicx}
\usepackage{amssymb,amsmath}
\usepackage{indentfirst}
\usepackage{pifont}
\usepackage{algorithmic}
\usepackage{algorithm}
\usepackage{cases}
\usepackage{soul}
\usepackage{booktabs,tabularx,hyperref}
\newcolumntype{Y}{>{\raggedright\arraybackslash}X} % 左对齐自动换行
\usepackage[table]{xcolor}
\usepackage{graphicx}
\IEEEoverridecommandlockouts
\makeatletter
\usepackage{times,verbatim,amsfonts,amsmath,color}
\usepackage{hyperref}
\makeatletter
\def\UrlAlphabet{%
      \do\a\do\b\do\c\do\d\do\e\do\f\do\g\do\h\do\i\do\j%
      \do\k\do\l\do\m\do\n\do\o\do\p\do\q\do\r\do\s\do\t%
      \do\u\do\v\do\w\do\x\do\y\do\z\do\A\do\B\do\C\do\D%
      \do\E\do\F\do\G\do\H\do\I\do\J\do\K\do\L\do\M\do\N%
      \do\O\do\P\do\Q\do\R\do\S\do\T\do\U\do\V\do\W\do\X%
      \do\Y\do\Z}
\def\UrlDigits{\do\1\do\2\do\3\do\4\do\5\do\6\do\7\do\8\do\9\do\0}
\g@addto@macro{\UrlBreaks}{\UrlOrds}
\g@addto@macro{\UrlBreaks}{\UrlAlphabet}
\g@addto@macro{\UrlBreaks}{\UrlDigits}
\makeatother

\usepackage[switch,pagewise]{lineno}

\begin{document}

\title{Agent Discovery in Internet of Agents: Challenges and Solutions}
%Capability Discovery in the Internet of Agents: Challenges and Solutions
% Agent Capability Discovery for Large-Scale Autonomous Collaboration: Principles, Challenges, and the Road Ahead
% Semantic Capability Discovery in the Internet of Agents: A Tutorial and Case Study
\author{
\IEEEauthorblockN{
Shaolong~Guo, 
Yuntao~Wang, 
Zhou~Su\IEEEauthorrefmark{1}, 
Yanghe~Pan,
Qinnan~Hu, and 
Tom H.~Luan}\\
\IEEEauthorblockA{
\IEEEauthorrefmark{0}School of Cyber Science and Engineering, Xi'an Jiaotong University, China 
\\ \IEEEauthorrefmark{2}Corresponding author: zhousu@ieee.org 
}}
\maketitle

\begin{abstract}
Rapid advances in large language models and agentic AI are driving the emergence of the Internet of Agents (IoA), a paradigm where billions of autonomous software and embodied agents interact, coordinate, and collaborate to accomplish complex tasks. A key prerequisite for such large-scale collaboration is agent capability discovery, where agents identify, advertise, and match one another's capabilities under dynamic tasks. Agent's capability in IoA is inherently heterogeneous and context-dependent, raising challenges in capability representation, scalable discovery, and long-term performance. To address these issues, this paper introduces a novel two-stage capability discovery framework. The first stage, \textit{autonomous capability announcement}, allows agents to credibly publish machine-interpretable descriptions of their abilities. The second stage, \textit{task-driven capability discovery}, enables context-aware search, ranking, and composition to locate and assemble suitable agents for specific tasks. Building on this framework, we propose a novel scheme that integrates semantic capability modeling, scalable and updatable indexing, and memory-enhanced continual discovery.
Simulation results demonstrate that our approach enhances discovery performance and scalability. Finally, we outline a research roadmap and highlight open problems and promising directions for future IoA.

\end{abstract}


\section{Introduction}\label{sec:introduction}
\IEEEPARstart{R}{apid} advances in large language models (LLMs) and agentic AI have spawned a diverse ecosystem of software and embodied agents. These agents power applications from conversational assistants to domestic and industrial robots and are increasingly endowed with sophisticated perception, reasoning, and action capabilities. As estimated \cite{grand2030AIagent}, the global AI agents market is roughly USD 5.4 billion in 2024, rising to about USD 50.3 billion by 2030. As these agents proliferate, large populations need to interconnect, communicate, and coordinate to tackle complex, multi-stage tasks that exceed the capacity of any single agent. This trend motivates the Internet of Agents (IoA): a vision of networked, autonomous agents \cite{chen2025ioa,yang2025agentic}, ranging from LLM-driven software entities to embodied robots, that interact and cooperate to accomplish tasks. Unlike traditional Internet of Things (IoT)’s passive data collection, the IoA envisions a decentralized, agent-centric ecosystem in which autonomous agents perceive their environment, make local decisions, and form adaptive collaborations, enabling a new era of distributed, resilient, and scalable intelligence.

A key prerequisite for large-scale inter-agent collaboration in the IoA is agent discovery, which involves identifying suitable agents based on their capabilities (e.g., available tools and knowledge) and the operational conditions under which those abilities can be utilized.
In the IoA, agent capability discovery is characterized by:
\begin{itemize}
    \item \textit{Heterogeneity}: Agents differ significantly in embodiment (e.g., software or hardware), functional roles, operational constraints (e.g., latency and energy), and toolchains \cite{yang2025autohma}. Their capabilities are structurally diverse and often not directly comparable.
    \item \textit{Autonomy}: Agents may advertise and update their capabilities and can autonomously discover suitable peers \cite{wang2025internet}, thereby favoring decentralized peer-to-peer discovery.
    \item \textit{Contextual dependence}: An agent's capability depends on task context \cite{xu2024when,jiang2024large} and internal state (e.g., an agent useful for perception in one scenario may be unsuitable for manipulation in another). As such, agent capabilities are highly time-varying and context-sensitive.
\end{itemize}
These characteristics raise a central question for IoA systems: \textit{how can agents efficiently discover, evaluate, and synchronize others’ capabilities across a large-scale, heterogeneous, and dynamic network?}

To meet these demands, we introduce a novel two-stage capability discovery framework for the IoA.  
\begin{itemize}
\item \textit{Stage 1: autonomous capability announcement}, where each agent publishes and synchronizes a machine-interpretable profile that describes its capabilities (e.g., built-in skills and tool access) and operating constraints in a decentralized and verifiable manner, thereby answering ``What can I offer?".  
\item \textit{Stage 2: task-oriented capability discovery}, where a requesting agent formulates its intent as a semantic query and performs context-aware search, ranking, and composition over the registry to identify suitable collaborators and assemble them into a workflow, thereby answering ``What do I need?".  
\end{itemize}
While this framework facilitates inter-agent collaboration, several key challenges remain. 
First, it remains challenging to build a unified and semantically expressive capability model to represent heterogeneous agents with diverse types and continuously evolving functionalities (e.g., tools and skills). Second, capability discovery faces the 
scalability challenge with massive agent populations while preserving semantic precision. Finally, in dynamic environments where agents frequently join, leave, or update their capabilities, sustaining long-term and consistent discovery performance remains difficult.

To address these limitations, we propose a semantic-driven capability discovery scheme that integrates pre-trained language models with scalable indexing mechanisms. It comprises three key phases: (i) a semantic profiling module that leverages language models to embed structured agent descriptions into a unified latent space, (ii) a compact and incrementally updatable index that supports efficient large-scale retrieval, and (iii) a memory-enhanced continual discovery component that retains historical knowledge while adapting to evolving agent capabilities. Furthermore, extensive simulations demonstrate that our approach improves both discovery performance and scalability. Finally, we outline a future research roadmap that identifies open challenges and promising directions.  

The remainder of this paper is organized as follows. Section~\ref{sec:architecture} introduces the IoA architecture and its capability management process. Section~\ref{sec:challenges} discusses the core technical challenges. Section~\ref{sec:casestudy} presents a case study of agent discovery scheme to address these challenges. Section~\ref{sec:futurework} highlights open research opportunities, and Section~\ref{sec:conclusion} concludes the paper.

% 第二章：IoA 架构与能力管理流程
\section{Overview of IoA Architecture and Agent Capability Discovery}\label{sec:architecture}
This section introduces the foundational architecture of IoA and elaborates on the workflow of capability discovery.

\subsection{IoA Architecture for Capability-driven Autonomous Collaboration}
\begin{figure}
    \centering\setlength{\abovecaptionskip}{-0.05cm}
    \includegraphics[width=0.8\linewidth]{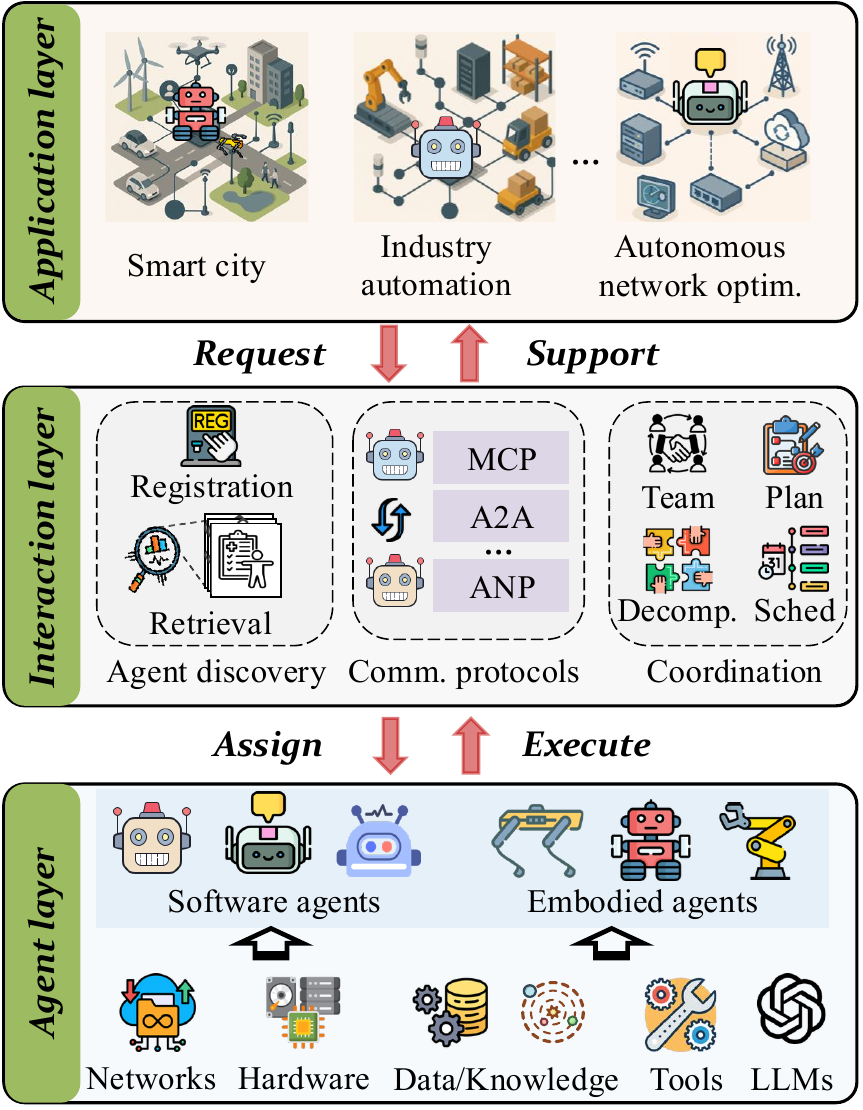}
    \caption{IoA Architecture for Capability-Driven Autonomous Collaboration.}
    \label{fig:arch}\vspace{-3mm}
\end{figure}

The IoA interconnects heterogeneous agents with diverse capabilities that can actively sense, reason, make decisions, and collaborate to accomplish tasks autonomously. To enable such large-scale, capability-oriented collaboration, the IoA adopts a three-layer architecture, as illustrated in Fig.~\ref{fig:arch}.

% Unlike conventional IoT, which mainly treats devices as passive data sources or sinks, the IoA features agents that actively sense, reason, decide, and cooperate to accomplish tasks in a fully autonomous manner. To enable large-scale, capability-driven collaboration, the IoA architecture is structured into three essential layers, as illustrated in Fig.~\ref{fig:architecture}.

\begin{itemize}
  \item \emph{Agent Layer.} This layer hosts a diverse set of autonomous agents \cite{yang2025autohma}, ranging from software-based entities (e.g., LLM-powered assistants) to embodied agents (e.g., robots and drones). As shown in Fig.~\ref{fig:arch}, these agents are supported by the IoA infrastructure, including communication resources, computing platforms (GPUs/CPUs/NPUs), shared data/knowledge, and tool APIs. Their capabilities in perception, reasoning, memory, planning, and control vary according to their form, operating environment, and available resources.
 Collectively, these heterogeneous agents form the foundation of scalable, distributed IoA services.

\item \emph{Interaction Layer.}  
Serving as the connective hub of the IoA, this layer enables agent discovery, agent-native communication, and autonomous coordination. It is built on three core pillars. 
First, \emph{agent discovery} supports registration and retrieval of agent profiles, allowing agents with the required capabilities to be quickly discovered on demand.  
Second, \emph{communication protocols} such as Agent-to-Agent (A2A) \cite{A2A} and Agent Network Protocol (ANP) \cite{ANP} provide standardized, agent-native interfaces that ensure secure interoperability across heterogeneous platforms, supporting context exchange, tool access, and peer-to-peer collaboration.  
Third, \emph{coordination mechanisms} enable agents to self-organize into teams, perform planning, decompose tasks, and schedule actions to form executable workflows, enabling collaboration autonomously.  
Together, these pillars constitute the operational backbone of the IoA, transforming isolated agent capabilities into collaborative intelligence.

  \item \emph{Application Layer.} 
This layer defines the interface where IoA services are delivered to end users and system operators. It enables agents to work toward shared goals by organizing their capabilities into task-driven workflows that support real-world applications across domains.
As shown in Fig.~\ref{fig:arch}, representative scenarios include: \emph{i) smart cities}, where traffic controllers, drones, and emergency responders form temporary teams to detect incidents and coordinate multi-modal responses in real time \cite{kalyuzhnaya2025llm,wang2025internet}; \emph{ii) industry automation}, where on-site agents (e.g., robotic arms) collaborate with external partners such as suppliers and UAV couriers to adapt production to real-time conditions \cite{wang2025internet}; and \emph{iii) autonomous network optimization}, where agent-based controllers in next-generation wireless systems (e.g., 6G) manage spectrum, optimize routing, and orchestrate edge resources for low-latency, energy-efficient communication \cite{chen2024enabling,jiang2024large}.

\end{itemize}

\begin{figure}
    \centering\setlength{\abovecaptionskip}{-0.05cm}
    \includegraphics[width=0.9\linewidth]{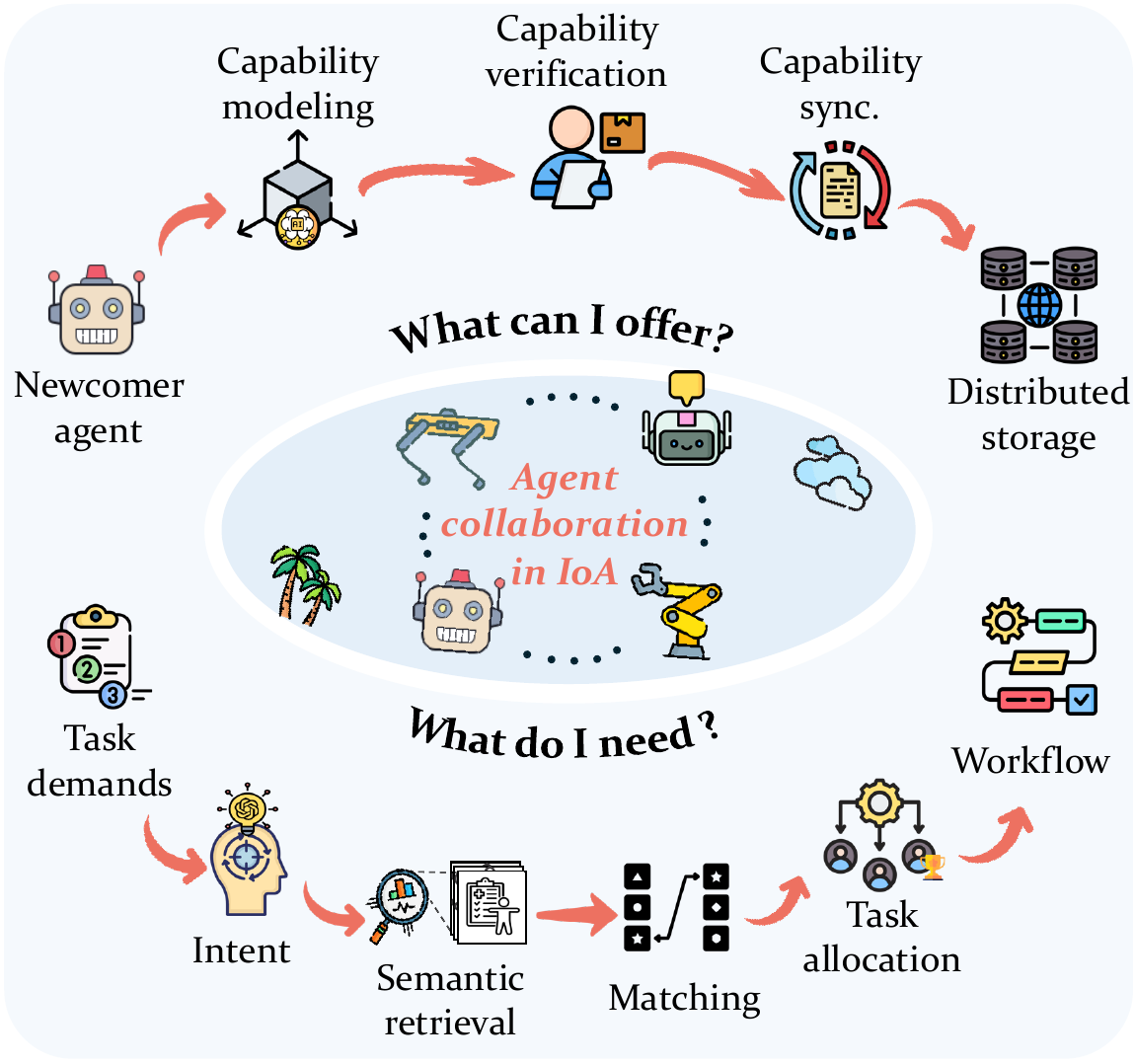}
    \caption{Overview of agent discovery in IoA. Each agent advertises its capabilities (i.e., \textit{what can I offer}) and discovers others based on task demands (i.e., \textit{what do I need}), enabling autonomous collaboration.}
    \label{fig:workflow}\vspace{-3mm}
\end{figure}

\newcolumntype{C}[1]{>{\centering\arraybackslash}m{#1}}
\begin{table*}[!t]
\centering
\setlength{\abovecaptionskip}{0cm}
\renewcommand{\arraystretch}{1.05} % 调小行距让表格更紧凑
\caption{Representative Discovery Mechanisms for the IoA}\label{tab:discovery_mechanisms}
\footnotesize
\resizebox{\textwidth}{!}{%
\begin{tabular}{C{1.85cm}|C{2.6cm}|C{2.4cm}|C{2.5cm}|C{1.4 cm}|C{2.0cm}|C{1.6cm}}
\toprule
\textbf{Solution} & \textbf{Description} & \textbf{Registration} & \textbf{Discovery} & \textbf{Metadata (form)} & \textbf{Features} & \textbf{Use cases} \\ \hline
\textbf{MCP Registry\footnotemark[1]} & Open catalog and API for publicly available MCP servers & Publisher submits entry; verified via GitHub or domain proof & Clients query Registry API to list/filter servers & server.json (MCP manifest) & Centralized; verified namespaces; open ecosystem & GitHub; Anthropic; Cloudflare \\ \hline
\textbf{A2A Agent Cards\footnotemark[2]} & Self-hosted JSON card for agent capabilities & Self-publishing at standard domain location & Fetch Agent Card (e.g., via Well-Known URI) & Agent Card (JSON) & Decentralized; interoperable; flexible & Google A2A codelabs \\ \hline
\textbf{Microsoft Entra Agent ID\footnotemark[3]} & Enterprise-level ID to track, manage, and secure agents & Agents from Azure AI Foundry/ Copilot Studio appear automatically & Directory-based discovery and governance & Agent identity record (Entra object) & Lifecycle governance; conditional access; enterprise integration & Azure AI Foundry; Copilot Studio \\ \hline
\textbf{Synergetics AgentConnect\footnotemark[4]} & Decentralized secure connectivity for agents and external Apps & Agents onboard via AgentID/ AgentWallet and register in AgentRegistry & Resolve agent via AgentRegistry, then establish a WalletConnect session & — & Secure exchange; payment integration; dApps/Web2 connectivity & AgentWorks suite (e.g., AgentWallet) \\ \hline
\textbf{NANDA AgentFacts\footnotemark[5]} & TTL-based endpoint resolution and capability verification & Publish AgentFacts and registered in NANDA Index & Two-stage: 1) lean index; 2) FactsURL/ PrivateFactsURL & AgentFacts (JSON-LD, verifiable credentials) & Lightweight, verifiable, and privacy-preserving discovery & PoC \\ \hline
\textbf{Agent Name Service (ANS)\footnotemark[6]} & DNS-inspired and PKI-anchored framework for AI agent discovery & Agent submits request, RA validates, and CA issues PKI certificate & Capability-aware lookup via protocol adapter layer & ANS registry record (JSON Schema) & Protocol-agnostic registry; trusted; multi-standard support & PoC \\
\bottomrule
\end{tabular}%
}
\end{table*}

\subsection{Workflow of Agent Discovery in IoA}
To enable autonomous collaboration in large-scale IoA, agents should not only advertise their \emph{built-in skills} but also seek suitable collaborators in a dynamic and decentralized environment. Agent discovery mechanisms aim to address two key questions: ``\emph{What can I offer?}" and ``\emph{What do I need?}" through two integrated phases, i.e., \emph{capability announcement} and \emph{capability discovery}, as illustrated in Fig.~\ref{fig:workflow}.

\subsubsection{Distributed capability announcement, evaluation, and synchronization}
This process ensures that every agent can quickly expose its capabilities to the IoA in an on-demand and verifiable manner, as well as synchronize others' capabilities.

\begin{itemize}
\item \emph{Capability modeling:} Agents describe their skills, ranging from deployed AI models and API/tool access to embodied interaction abilities (e.g., sensing and mobility), in a structured, machine-readable format. The description encompasses both functional capabilities and non-functional attributes (e.g., latency tolerance, energy constraints) as well as relevant context.

\item \emph{Capability verification:} To establish credibility and foster trust, declared capabilities by agents are validated through standardized benchmarks, sandbox trials, or peer endorsements. Such a selective verification process balances efficient validation with the reliability of claimed capabilities.

\item \emph{Capability synchronization:} Given the IoA’s highly dynamic nature, agents proactively disseminate updates on capability changes (e.g., newly acquired functions or degraded performance) using event-driven or state-triggered mechanisms~\cite{wang2025internet}. Such an approach ensures the capability registry remains up-to-date with modest overhead.

\item \emph{Distributed registry storage:} All capability metadata and associated credibility metrics can be maintained in a decentralized registry (e.g., distributed hash tables or gossip-based indexing), supporting scalable, fault-tolerant, and censorship-resistant capability discovery.
\end{itemize}

\footnotetext[1]{\url{https://github.com/modelcontextprotocol/registry}} % MCP Registry (official MCP registry repo)
\footnotetext[2]{\url{https://a2a-protocol.org/latest/topics/agent-discovery/}}                 % A2A Protocol (official docs)
\footnotetext[3]{\url{https://techcommunity.microsoft.com/blog/microsoft-entra-blog/announcing-microsoft-entra-agent-id-secure-and-manage-your-ai-agents/3827392}} % Microsoft Entra Agent ID (official announcement)
\footnotetext[4]{\url{https://synergetics.ai/platform/agentconnect/}}     % AgentConnect (official vendor page)
\footnotetext[5]{\url{https://nanda.media.mit.edu/}}        % NANDA / AgentFacts (official MIT page)
\footnotetext[6]{\url{https://genai.owasp.org/resource/agent-name-service-ans-for-secure-al-agent-discovery-v1-0/}} % ANS (official OWASP GenAI resource)

\subsubsection{Task-oriented Capability Discovery}
This stage enables agents to quickly identify and engage with the most suitable collaborators within the IoA for a given task.

\begin{itemize}
\item \emph{Intent understanding:} 
Task descriptions, whether expressed in natural language or formal protocol, are translated into semantic representations that precisely capture both the functional objectives and operational constraints, forming a precise query for agents matching with suitable capability.

\item \emph{Semantic retrieval and candidate matching:} 
The task representation can be matched against the capability registry using a multi-criteria ranking framework that integrates semantic similarity, credibility scores, contextual relevance, real-time availability, etc. This process generates a prioritized candidate list of agents most likely to fulfill the task demands with high fidelity.

\item \emph{Task allocation and workflow orchestration:} 
Once suitable agents are identified, the IoA coordinates task allocation and orchestrates collaborative workflows among them. Communication protocols provide a standardized interface for seamless information exchange and interoperability across heterogeneous agents. To ensure robust collaboration, flexible consensus and conflict-resolution mechanisms are integrated to resolve discrepancies in decision-making and ensure consistent task execution.
\end{itemize}

Table~I summarizes the representative agent discovery mechanisms for the IoA deployments.

% 第三章：核心挑战&Solutions
% todo: 虽然诸如 A2A 协议已经设置了 agent card 来回答了一个基本问题：“这个代理可以做什么，我该如何与它交谈？但发现需要的不仅仅是名片，还需要一个系统来找到适合您需求的名片。虽然代理卡提供了基础，但构建动态发现系统需要 A2A 有意将三个额外的组件留给实施者：1. Agent Registration; 2. Capability-Based Discovery: Agent Naming Service; 3. Resolution and Policy Enforcement: Agent Gateway

\section{Key Challenges to Agent Capability Discovery in IoA}\label{sec:challenges}
% This section outlines the primary technical challenges in IoA capability discovery and provides an overview of cutting-edge solutions targeting the core challenges.
% This section highlights the key technical challenges in IoA capability discovery and outlines frontier solutions designed to overcome them.

% \subsection{Key Challenges}
\subsection{Low Expressivity in Heterogeneous Capability Modeling}\label{sec:expression}
% 核心问题（低表达性）：IoA 生态中有软件型 LLM Agent、API Agent、机器人等多种异构实体，能力动态生成、环境多模态，现有基于静态本体（ontology）或手工模板的建模无法高效表达复杂语义、上下文约束以及多模态交互信息。本质痛点：针对多种异构实体，缺乏统一、动态、高语义表达力的能力表示标准，导致能力无法精确描述、难以互操作，阻碍能力发现和复用。
IoA ecosystems encompass a wide spectrum of heterogeneous agents, ranging from software-based LLM agents to physically embodied robots. A fundamental challenge lies in representing these diverse and dynamic capabilities with high semantic expressivity and computational interpretability. Existing modeling schemes, either based on static ontologies or handcrafted templates, are inadequate for capturing the complex behaviors of heterogeneous agents that continuously evolve and generate new capabilities on the fly. Moreover, embodied agents operate in multimodal environments and require fine-grained semantic models to express context, constraints, and sensor-actuator dependencies. Without a unified and expressive capability representation, agents cannot accurately advertise their potential nor be effectively understood by other agents. This hampers collaboration and significantly reduces the efficiency of capability discovery. Hence, it is imperative to design a semantic modeling paradigm that supports abstraction-to-execution mapping, accommodates dynamic generation, and enables multimodal semantic translation across heterogeneous agents.

\subsection{Limited Scalability in Agent Capability Retrieval}\label{sec:matching}
% 核心问题（低可扩展性）：IoA 系统规模巨大，百万级 Agent、动态能力库，任务意图常用自然语言表达，检索必须在高模糊度、低延迟条件下完成。传统 sparse (e.g., keyword)和dense检索无法满足语义精度和实时性。本质痛点：如何在大规模、动态索引下实现低延迟、高语义精度的能力检索，尤其是面向模糊、高抽象意图。
As IoA ecosystems scale to support large numbers of agents with evolving capabilities, agent discovery faces significant scalability challenges. In particular, matching abstract or fuzzy task requests to the most relevant agents becomes increasingly difficult as the pool of candidates expands.
Traditional retrieval mechanisms struggle to meet these demands. Sparse retrieval techniques (e.g., keyword matching) \cite{robertson2009probabilistic} fail to capture semantic meaning, while dense retrieval approaches \cite{karpukhin2020dense} rely on high-dimensional vector indexes that impose substantial memory overhead and retrieval latency as the number of agents increases.
Such latency not only degrades overall IoA efficiency but also impairs time-sensitive applications (e.g., disaster relief). To address these challenges, there is a pressing need for efficient and semantics-aware discovery frameworks capable of supporting accurate retrieval under large-scale, high-concurrency conditions.

\subsection{Inconsistency in Long-term Agent Discovery in Dynamic IoA}\label{sec:dynamic}
% 核心问题（不一致同步）：Agent 能力随资源状态、环境、学习过程动态变化，基于周期更新或集中轮询易造成信息陈旧。本质痛点：如何在分布式环境中实现低开销、强一致或最终一致的能力状态同步，保障发现与调度基于最新状态，防止 stale data 带来的协作风险。
In dynamic IoA environments, agents may frequently join or leave the ecosystem, and their capabilities evolve continuously due to changing context, available resources, and operational status. A key challenge lies in maintaining an accurate and up-to-date view of agents' capabilities across the system. Relying on periodic updates or centralized polling often results in stale information, which may lead to unsuitable agent selection and degraded coordination quality. Moreover, the discovery mechanism should incorporate newly registered agents in real-time while ensuring that existing agents remain retrievable as the system evolves. This demands not only real-time responsiveness but also long-term retrieval performance without forgetting valuable knowledge over time. Thus, it is critical to design a lightweight and sustained discovery mechanism for continual cooperation in dynamic IoA.

\section{Case Study: Semantic-Driven Capability Discovery in IoA}\label{sec:casestudy}
\subsection{Design Overview}\label{method0}
To address these challenges mentioned in Sect.~\ref{sec:challenges}, as shown in Fig.~3, we propose a semantic-driven capability discovery framework tailored for IoA.

\begin{itemize}
    \item \textit{Phase 1: Semantic Agent Profiling (Sect.~\ref{method1}).} 
    {Leveraging the semantic understanding and reasoning capabilities of language models, we encode agent profiles into unified semantic representations, where functionally similar agents are positioned closer together. This semantic space provides the foundation for subsequent indexing and discovery.}

    \item \textit{Phase 2: Scalable Agent Indexing (Sect.~\ref{method2}).} 
    {To support large-scale deployment, we design an efficient indexing mechanism that partitions high-dimensional representations into subspaces, performs clustering within each subspace, and assigns compact agent codes based on representative vectors of the clusters. These codes serve as lightweight yet stable IDs, allowing new agents to be integrated without altering existing assignments.}

    \item \textit{Phase 3: Continual Agent Discovery (Sect.~\ref{method3}).} 
    {Building on the indexed codes, we train the retrieval model with a memory-preserving mechanism. By replaying representative historical agents alongside new arrivals, the model learns to map task queries directly to agent codes while mitigating forgetting, thus sustaining accurate and resilient capability discovery over time.}
\end{itemize}
Together, these phases form a semantic-driven pipeline that enables accurate, scalable, and lifelong agent discovery in dynamic IoA environments.

\begin{figure}
    \centering\setlength{\abovecaptionskip}{-0.05cm}
    \includegraphics[width=0.95\linewidth]{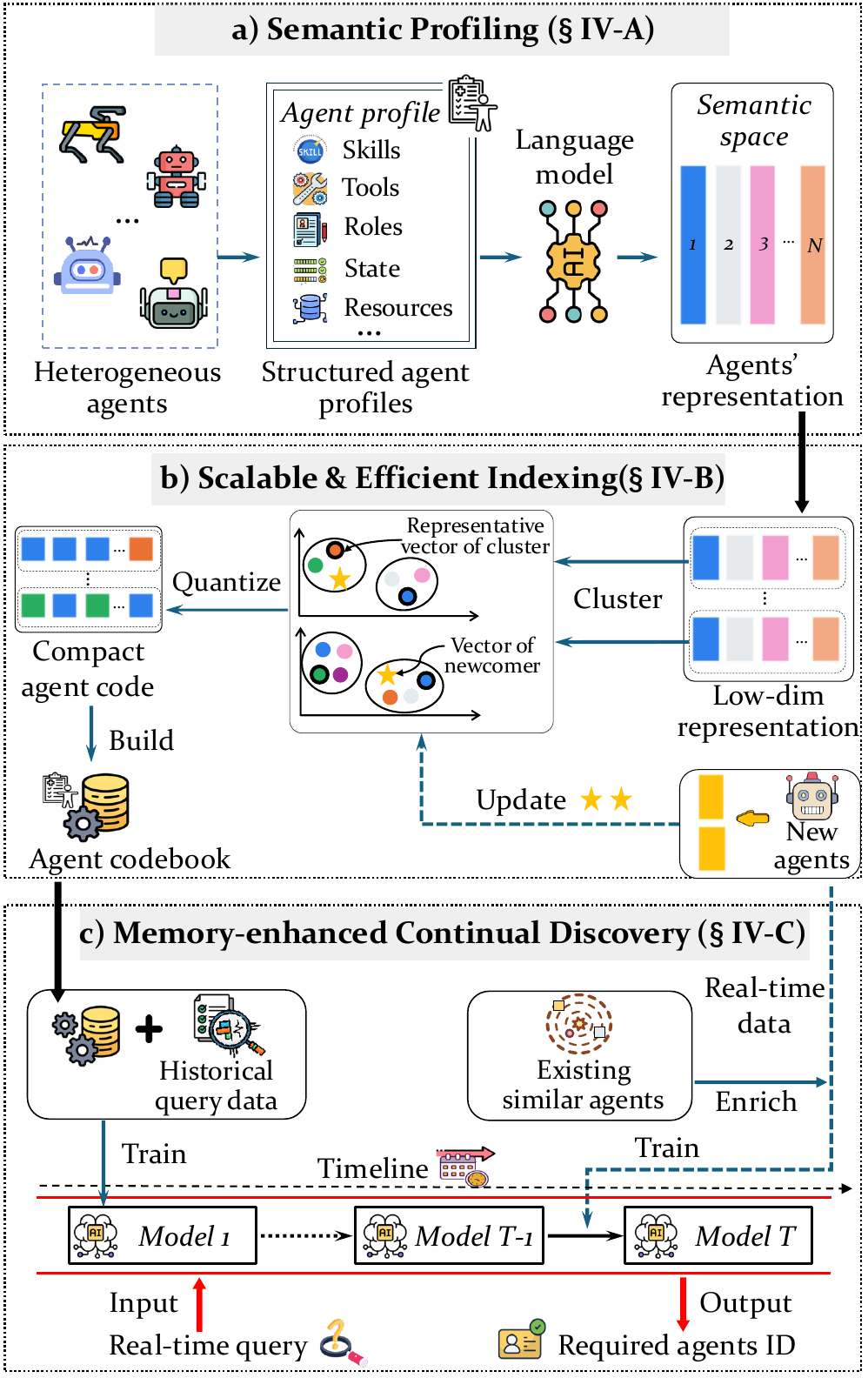}
    \caption{Illustration of semantic-driven capability discovery solutions in IoA, including: (a) language model-powered semantic profiling of agents, (b) scalable and efficient indexing, and (c) memory-enhanced continual discovery.}
    \label{fig:case}
\end{figure}

\subsection{Language Model-powered Semantic Profiling of Agents}\label{method1}
Enabling accurate and flexible agent discovery in IoA requires more than static capability tags. As agents become increasingly heterogeneous, multimodal, and language-capable, it becomes essential to represent their functional roles and interaction traits in a semantically meaningful way. To address this, as shown in Fig.~\ref{fig:case} (a), we develop a language model-powered semantic profiling mechanism that transforms agent metadata into expressive, machine-interpretable embeddings for downstream indexing and retrieval, with the following two stages.

\textit{Stage 1: Structured Agent Profiling.}  
Each agent autonomously constructs a structured profile that includes three core dimensions:
\begin{itemize}
    \item \textit{Skills and tools:} Executable actions or APIs exposed by the agent, such as ``route planning", ``sentiment classification", or ``environmental monitoring".
    \item \textit{Roles and expertise:} High-level operational functions, including planner, assistant, sensor, or translator.
    \item \textit{State and constraints:} Contextual factors such as memory capacity, latency tolerance, hardware location (e.g., cloud or edge), and current load.
\end{itemize}

%\textcolor{blue}{\textit{Stage 2: Semantic Representation via Language Models.}  
{\textit{Stage 2: Semantic Representation via Language Models.}  
The structured agent profile is then serialized into text and encoded using a pre-trained language model (e.g., BERT or DistilRoBERTa) to generate a dense semantic representation. These representations reside in a unified latent space where agents with similar functions, but possibly different lexical descriptions, are positioned closely. For instance, agents described as ``path planning" and ``route optimization" are likely to be matched, even without a shared vocabulary. Such semantic encoding enables robust cross-domain discovery and improves generalization across heterogeneous agents, while also being maintained in a unified repository as the basis for subsequent indexing and scalable discovery.}

%%%

\subsection{Scalable \& Efficient Indexing for Large-scale IoA}\label{method2}

To enable end-to-end agent retrieval, each agent should be assigned a compact semantic-rich identifier (ID) that can be directly generated and matched during discovery. However, using high-dimensional embeddings as IDs is impractical, as they incur high memory and computational costs as agents evolve. These challenges are further amplified in large-scale and dynamic IoA environments, where discovery demands an efficient and adaptive indexing mechanism. As illustrated in Fig.~\ref{fig:case} (b), we propose a scalable indexing framework that compresses agent embeddings into discrete codes. The resulting codes substantially reduce storage and retrieval overhead, enable fast and semantically consistent identification, and support continual updates without imposing heavy system costs.

\textit{Stage 1: Compact Codebook Construction.} We construct a lightweight codebook following a product quantization-inspired approach. Each high-dimensional agent embedding is first partitioned into multiple subspaces. Within each subspace, clustering is applied to identify representative vectors that serve as semantic anchors. Each agent is then assigned a discrete code by mapping its sub-vectors to the nearest anchors across subspaces. The resulting code acts as a lightweight yet discriminative ID that preserves functional similarity among agents and supports efficient downstream retrieval.

\textit{Stage 2: Incremental Index Maintenance.}  
When new agents join or existing agents update their capabilities, their embeddings are also partitioned under the subspaces. If the resulting sub-vectors align well with existing clusters, they are directly encoded using the nearest anchors. Otherwise, local refinements are performed in the affected subspaces to capture the new semantics while leaving existing assignments unchanged. As such, this strategy allows the index to evolve gracefully with the agent population, maintaining retrieval efficiency and adaptability in dynamic IoA environments.

\subsection{Memory-enhanced Continual Discovery}\label{method3}
Beyond scalable indexing, the retrieval model should be continually trained to map task queries to agent IDs. As the agent population evolves, the retrieval model should seamlessly integrate newly arriving agents without forgetting past knowledge of existing ones. As illustrated in Fig.~\ref{fig:case} (c), we introduce a memory-enhanced continual discovery framework that combines selective replay with simple stability controls.

\textit{Stage 1: End-to-end Training with Knowledge Replay.} 
The retrieval model is trained in an end-to-end manner to generate required agents' IDs from task queries. To prevent forgetting, a lightweight memory buffer stores representative records of previously registered agents. During each update, existing semantically similar samples are replayed alongside new agent data, ensuring that long-standing agents remain discoverable as the model learns to encode new capabilities.

\textit{Stage 2: Stability-constrained Updates.} 
To further mitigate semantic drift, stability constraints are applied during model updates. 
Embedding dimensions most critical for retrieving historical agents are selectively protected, while less sensitive parameters are allowed to adapt to new semantics. This controlled update strategy enables the embedding space to evolve incrementally, capturing emerging functionalities without compromising retrieval fidelity for established agents.

\subsection{Performance Evaluation}
We conduct simulations to validate the effectiveness and scalability of the proposed capability-aware agent discovery framework in IoA environments. The language model is implemented using pre-trained BERT models to generate dense semantic embeddings from structured agent capability descriptions. 
For evaluation, our framework is deployed on a server equipped with dual Intel Xeon Platinum 8378C CPUs, 512 GB of RAM, and two NVIDIA A100 GPUs (80 GB memory each). We compare our approach with two widely used baselines: (i) BM25-based sparse retrieval \cite{robertson2009probabilistic}, and (ii) dense retrieval \cite{karpukhin2020dense} using a standard dual-encoder setup.

\begin{figure}
\setlength{\abovecaptionskip}{-0.01cm}
    \centering
    \includegraphics[width=0.85\linewidth]{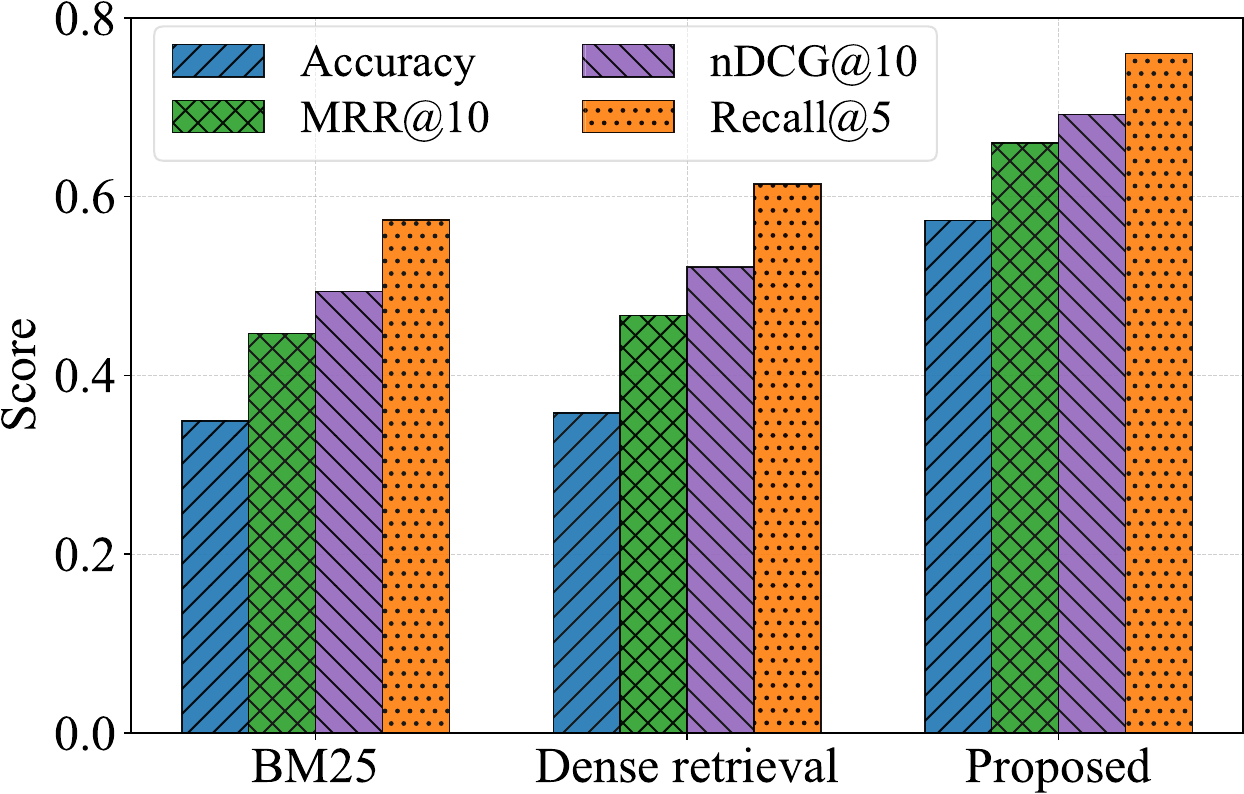}
    \caption{Comparison of agent discovery performance between ours and benchmarks.}
    \label{fig:discovery_performance}\vspace{-2mm}
\end{figure}

\begin{figure}
\setlength{\abovecaptionskip}{-0.01cm}
    \centering
    \includegraphics[width=0.85\linewidth]{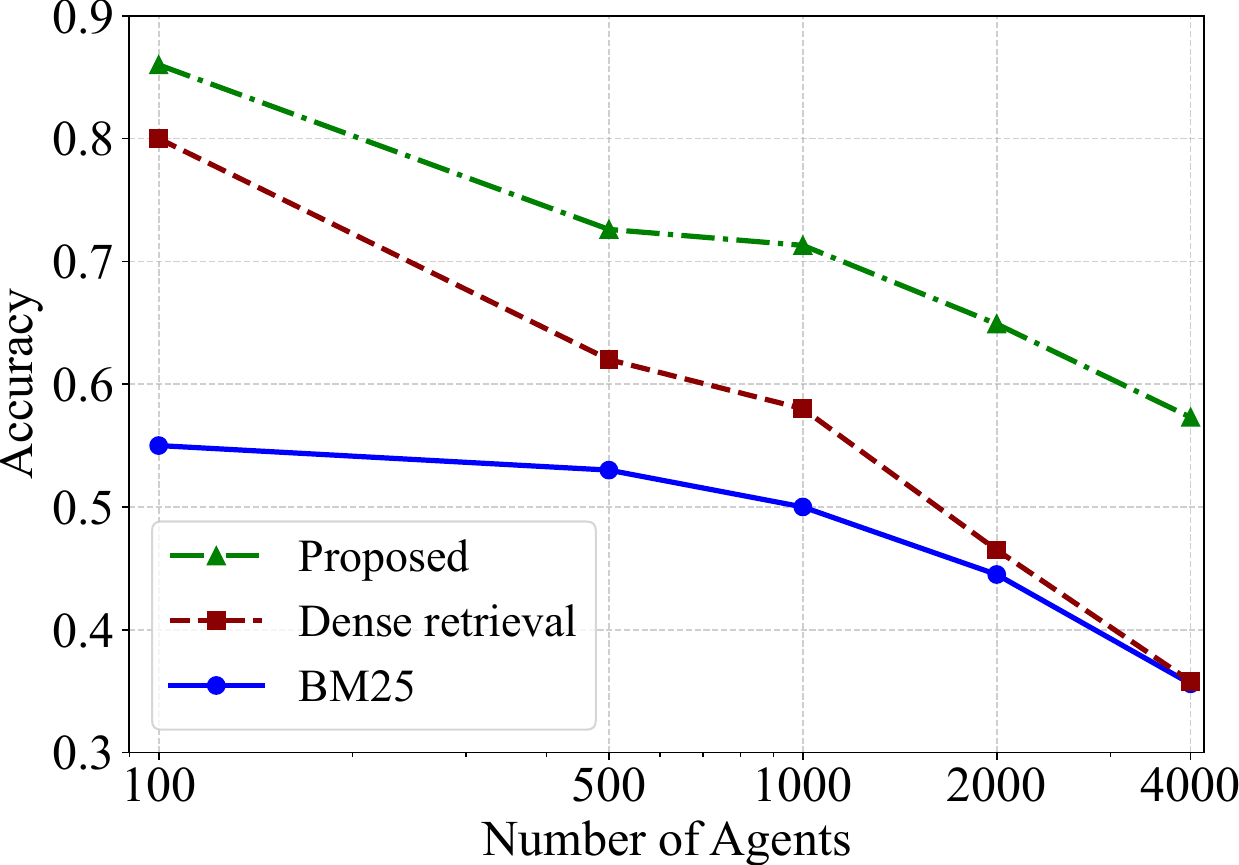}
    \caption{Comparison of discovery performance in three schemes under varying numbers of agents.}
    \label{fig:agents_vs_accuracy}\vspace{-3mm}
\end{figure}

We first evaluate the agent discovery performance across different schemes. As depicted in Fig.~\ref{fig:discovery_performance}, the proposed approach consistently outperforms both baselines across all metrics, including Accuracy, MRR@10, nDCG@10, and Recall@5. Notably, our approach achieves a Recall@5 of 0.76, representing a significant improvement of over 24\% compared to the best-performing baseline. The improvement in ranking-based metrics further demonstrates the effectiveness of our semantic representation and indexing approach in capturing functional similarity among diverse agents.

Next, we assess the scalability of our discovery framework as the number of involved agents increases. As illustrated in Fig.~\ref{fig:agents_vs_accuracy}, both BM25 and dense retrieval suffer substantial performance degradation as the agent pool grows. In contrast, the proposed approach maintains high discovery accuracy even with thousands of agents, thanks to the compact and adaptive indexing design. With 4,000 agents, our approach still achieves a top-1 accuracy of 0.58, compared to 0.35 and 0.36 for BM25 and dense retrieval, respectively. These results demonstrate the scalability of our approach for large-scale IoA scenarios.

% 第六章：未来工作与展望
\section{Future Research Directions}\label{sec:futurework}
This section discusses open problems that remain to be investigated in the field of agent discovery in IoA.

\subsection{Cross-domain Collaborative Agent Capability Discovery}
Future IoA ecosystems will connect agents from vastly different domains, such as medical diagnostic agents and industrial robots. A primary challenge is achieving cross-domain knowledge interoperability \cite{niu2024comprehensive} and discovering agents whose individual capabilities can be creatively combined to accomplish complex tasks. The true value of IoA emerges when agents can be composed in unforeseen ways to solve emergent problems that no single agent could tackle alone.
One promising direction is to explore the use of graph neural networks to capture and analyze the intricate relationships between agent capabilities, enabling the discovery of latent collaborative patterns and optimal capability compositions. Another emerging area is meta-learning for agent composition, where the system learns to quickly adapt and form new agent teams based on a few examples of successful collaborations, greatly enhancing the IoA's agility and autonomy.

\subsection{Secure and Privacy-preserving Agent Discovery}
%As the IoA scales up and extends into various sensitive domains, 
Ensuring security and privacy preservation in agent capability discovery is essential for building a trustworthy IoA ecosystem. A major concern is that agents may inadvertently expose sensitive information (e.g., physical location or permissions to access specific services or data) during the discovery process, leading to privacy leakage, profiling, or targeted attacks \cite{syros2025saga}.
To address this, future IoA systems should support credible capability verification without revealing sensitive details. A promising direction is to integrate blockchain with zero-knowledge proofs, enabling agents to generate verifiable capability credentials while maintaining confidentiality. Complementary techniques such as differential privacy and federated learning can further safeguard agent data and models throughout the discovery and collaboration lifecycle.
Another challenge is to design privacy-preserving mechanisms that maintain data confidentiality during system-wide deployment without sacrificing discovery accuracy and computational efficiency.

\subsection{Economic Models for Sustainable IoA Ecosystems}
For the IoA to thrive as a self-sustaining and vibrant ecosystem, it is imperative to design fair and efficient economic models that encourage agents to share their capabilities and contribute to collective tasks actively. This involves establishing transparent mechanisms for value exchange that accurately quantify the impact and effort of an agent's contribution. Without proper incentives, agents may hoard valuable capabilities or be unwilling to participate in complex collaborations \cite{wang2025internet,yang2025agentic}, thus hindering the construction of healthy IoA ecosystems. Future research will delve into the application of tokenomics, contribution-aware incentive schemes, and deterrent mechanisms for agent collaboration. Moreover, embedding reputation systems can help reinforce long-term cooperative behaviors and filter unreliable agents, thereby fostering trust and self-regulation within IoA. Reputation systems can also be effectively implemented via smart contracts to enable automated and trustless value transfers, as well as enforce agreements between collaborators. %Thereby, it ensures that agents' contributions are appropriately recognized and fostering a self-regulating marketplace for agent capabilities.

\section{Conclusion}\label{sec:conclusion}
Capability discovery is a cornerstone for enabling large-scale collaboration in the IoA. Yet, the heterogeneity of agents, along with their autonomy and task-dependent nature in capability discovery, introduces fundamental challenges in representation, scalability, and adaptability.
This paper has proposed a general capability discovery framework that integrates autonomous capability announcement with task-driven capability discovery to improve accuracy, efficiency, and decentralization in inter-agent collaboration. We have further analyzed key challenges in IoA deployments and devised corresponding solutions, including language-model-based capability modeling, scalable and incrementally updatable indexing, and memory-enhanced continual discovery. Simulation results validate the effectiveness of our approach in improving discovery performance and scalability. Looking ahead, we envision that the presented framework will spark new directions toward scalable, trustworthy, and adaptive agent collaboration in IoA ecosystems.
% This work concludes by reiterating the critical importance of capability discovery and management as the cornerstone of IoA, paving the way for a future of truly collaborative and intelligent networks.

% \bibliographystyle{IEEEtran}
\bibliographystyle{ieeetr} %overlaf use

\bibliography{ref.bib}

@article{wang2025internet,
  title={Internet of agents: Fundamentals, applications, and challenges},
  author={Wang, Yuntao and Guo, Shaolong and Pan, Yanghe and Su, Zhou and Chen, Fahao and Luan, Tom H and Li, Peng and Kang, Jiawen and Niyato, Dusit},
  journal={arXiv preprint arXiv:2505.07176},
  year={2025}
}

@ARTICLE{yang2025autohma,
  author={Yang, Tingting and Feng, Ping and Guo, Qixin and Zhang, Jindi and Zhang, Xiufeng and Ning, Jiahong and Wang, Xinghan and Mao, Zhongyang},
  journal={IEEE Trans. Cognit. Commun. Networking}, 
  title={{AutoHMA-LLM}: Efficient Task Coordination and Execution in Heterogeneous Multi-Agent Systems Using Hybrid Large Language Models}, 
  year={2025},
  volume={11},
  number={2},
  pages={987-998}}

@inproceedings{chen2025ioa,
title={Internet of Agents: Weaving a Web of Heterogeneous Agents for Collaborative Intelligence},
author={Weize Chen and Ziming You and Ran Li and Yitong Guan and Chen Qian and Chenyang Zhao and Cheng Yang and Ruobing Xie and Zhiyuan Liu and Maosong Sun},
booktitle={Proc. ICLR},
year={2025},
pages = {1--32}
}

@article{chen2024enabling,
  title={Enabling mobile {AI} agent in {6G} era: Architecture and key technologies},
  author={Chen, Ziqi and Sun, Qi and Li, Nan and Li, Xiang and Wang, Yang and others},
  journal={IEEE Network},
  volume={38},
  number={5},
  pages={66--75},
  year={2024},
  publisher={IEEE}
}

@article{kalyuzhnaya2025llm,
  title={LLM Agents for Smart City Management: Enhancing Decision Support Through Multi-Agent {AI} Systems.},
  author={Kalyuzhnaya, Anna and Mityagin, Sergey and Lutsenko, Elizaveta and Getmanov, Andrey and Aksenkin, Yaroslav and Fatkhiev, Kamil and Fedorin, Kirill and Nikitin, Nikolay O and Chichkova, Natalia and Vorona, Vladimir and others},
  journal={Smart Cities},
  volume={8},
  number={1},
  year={2025},
  pages={1-33}
}

@ARTICLE{xu2024when,
  author={Xu, Minrui and Niyato, Dusit and Kang, Jiawen and Xiong, Zehui and Mao, Shiwen and Han, Zhu and Kim, Dong In and Letaief, Khaled B.},
  journal={IEEE Wireless Commun.}, 
  title={When Large Language Model Agents Meet {6G} Networks: Perception, Grounding, and Alignment}, 
  year={2024},
  volume={31},
  number={6},
  pages={63-71}
}

@ARTICLE{jiang2024large,
  author={Jiang, Feibo and Peng, Yubo and Dong, Li and Wang, Kezhi and Yang, Kun and Pan, Cunhua and Niyato, Dusit and Dobre, Octavia A.},
  journal={IEEE Wireless Commun.}, 
  title={Large Language Model Enhanced Multi-Agent Systems for {6G} Communications}, 
  year={2024},
  volume={31},
  number={6},
  pages={48-55}

}

@misc{grand2030AIagent,
  title = {{AI} Agents Market Size, Share \& Trends | Industry Report 2030},
  author = {Grand View Research},
  howpublished = {\url {https://www.grandviewresearch.com/industry-analysis/ai-agents-market-report}},
  year = {2025},
  note = {Accessed: Sep. 1, 2025}
}

@misc{A2A,
  author = {Google},
  title = {Agent to Agent Protocol  ({A2A})},
  year = {2025},
  howpublished = {\url {https://a2a-protocol.org/latest/}},
  note = {Accessed: Sep. 1, 2025},
  urldate = {}
}

@misc{ANP,
  author = {},
  title = {Agent Network Protocol ({ANP})},
  year = {2024},
  howpublished = {\url {https://agentnetworkprotocol.com/en/}},
  note = {Accessed: Sep. 1, 2025},
  urldate = {}
}

@inproceedings{niu2024comprehensive,
  title     = {A Comprehensive Survey of Cross-Domain Policy Transfer for Embodied Agents},
  author    = {Niu, Haoyi and Hu, Jianming and Zhou, Guyue and Zhan, Xianyuan},
  booktitle = {Proc. IJCAI},
  pages     = {8197--8206},
  year      = {2024}
}

@article{yang2025agentic,
  title={Agentic web: Weaving the next web with {AI} agents},
  author={Yang, Yingxuan and Ma, Mulei and Huang, Yuxuan and Chai, Huacan and Gong, Chenyu and Geng, Haoran and Zhou, Yuanjian and Wen, Ying and Fang, Meng and Chen, Muhao and others},
  journal={arXiv preprint arXiv:2507.21206},
  year={2025}
}

@inproceedings{syros2025saga,
  title={{SAGA}: A security architecture for governing {AI} agentic systems},
  author={Syros, Georgios and Suri, Anshuman and Ginesin, Jacob and Nita-Rotaru, Cristina and Oprea, Alina},
  booktitle={Proc. NDSS},
  year={2026},
  pages = {1-23}
}

@inproceedings{karpukhin2020dense,
  title={Dense Passage Retrieval for Open-Domain Question Answering.},
  author={Karpukhin, Vladimir and Oguz, Barlas and Min, Sewon and Lewis, Patrick SH and Wu, Ledell and Edunov, Sergey and Chen, Danqi and Yih, Wen-tau},
  booktitle={Proc. EMNLP},
  pages={6769--6781},
  year={2020}
}

@article{robertson2009probabilistic,
  title={The probabilistic relevance framework: {BM25} and beyond},
  author={Robertson, Stephen and Zaragoza, Hugo and others},
  journal={Foundations and Trends{\textregistered} in Information Retrieval},
  volume={3},
  number={4},
  pages={333--389},
  year={2009}
}

\end{document}